\documentclass[aps,pra,floatfix,superscriptaddress,twocolumn]{revtex4}
\usepackage[dvips]{graphicx}
\usepackage{color}

\usepackage{longtable}
\usepackage{dcolumn}
\usepackage[dvips]{graphicx}
\usepackage{bm}
\usepackage{bbm}

\usepackage{times}
\usepackage{nicefrac}
\usepackage{amsmath}
\usepackage{amsfonts}
\usepackage{amssymb}
\usepackage{amsthm}

\newcolumntype{.}{D{x}{}{-1}}
\newcolumntype{w}[1]{D{.}{.}{#1}}

\newcommand{\balpha}{\bm{\alpha}}

\newcommand{\bfr}{{\bm {r}}}

\newcommand{\bfp}{{\bm {p}}}

\newcommand{\lbr}{\langle}
\newcommand{\rbr}{\rangle}

\newcommand{\Za}{Z\alpha}

\begin{document}

\title{Nonlinearities of King's plot and their dependence on nuclear radii}

\author{Robert A. M\"uller}
\affiliation{Physikalisch-Technische Bundesanstalt, D-38116 Braunschweig, Germany}
\affiliation{Technische Universit\"at Braunschweig, D-38106 Braunschweig, Germany}

\author{Vladimir A. Yerokhin}
\affiliation{Center for Advanced Studies, Peter the Great St. Petersburg State Polytechnical
University, 195251 St. Petersburg, Russia}

\author{Anton N. Artemyev}
\affiliation{Institut f\"ur Physik, Universit\"at Kassel, Kassel 34132, Germany}

\author{Andrey Surzhykov}
\affiliation{Physikalisch-Technische Bundesanstalt, D-38116 Braunschweig, Germany}
\affiliation{Technische Universit\"at Braunschweig, D-38106 Braunschweig, Germany}

\begin{abstract}
Investigations of isotope shifts of atomic spectral lines provide insights into nuclear
properties. Deviations from the linear dependence of the isotope shifts of two atomic
transitions on nuclear parameters, leading to a nonlinearity of the so-called King plot, are
actively studied as a possible way of searching for the new physics. In the present work we
calculate the King-plot nonlinearities originating from the Standard-Model atomic theory. The
calculation is performed both analytically, for a model example applicable for an arbitrary
atom, and numerically, for one-electron ions. It is demonstrated that the Standard-Model
predictions of the King-plot nonlinearities are hypersensitive to experimental errors of
nuclear charge radii. This effect significantly complicates identifications of possible
King-plot nonlinearities originating from the new physics.
\end{abstract}
\maketitle

{\em Introduction.---} Introduced by W.~H.~King in 1963 \cite{king:63}, the King plot has been
widely used as an extremely useful tool for interpreting results of the isotope-shifts atomic
experiments. After measuring energies of two transitions for four or more isotopes of the same
element, one can arrange the results as a King plot, which should be --  to a very high accuracy
-- linear. In this way the experimental results can be cross-checked without any further
theoretical input. Moreover, the two parameters of the linear plot can be interpreted in terms of
the mass and field shifts and separately compared with theoretical calculations \cite{king:84}.

Progress in quantum logic techniques and collinear spectroscopy achieved during the last years
resulted in a dramatic increase of the experimental accuracy. In particular, measurements of
optical-clock transitions were demonstrated on a few-Hertz precision level
\cite{knollmann:19,solaro:20}. Even higher accuracy can be achieved by using entangled states
\cite{manovitz:19} and coherent high-resolution optical spectroscopy \cite{micke:20}. Naturally,
a question arises if the King plot is going to stay linear at this new level of experimental
precision. The recent experiments in strontium and ytterbium ions \cite{miyake:19,counts:20} gave
first indications that the linearity is actually broken on the level of several standard
deviations.

It was recently demonstrated \cite{frigiuele:17,berengut:18,berengut:20} that the linearity of
the King plot can be used in searches for new physics, specifically, to constrain the coupling
strength of hypothetical new-physics boson fields to electrons and neutrons. It remained unclear,
however, whether any actually {\em observed} nonlinearity can be clearly interpreted as a
manifestation of new physics, since the Standard-Model theory can also produce effects that
slightly bend the King plot. Such nonlinear effects have been a subject of several recent studies
\cite{flambaum:18,flambaum:19,yerokhin:20:kingsplot}.

The King plot is defined so that its construction does not require any knowledge about the
nuclear charge radii; only the nuclear masses are involved. Any {\em prediction} of King-plot
nonlinearities, however, requires an experimental input in the form of nuclear radii. One might
assumed that in view of the extreme smallness of the nonlinearities, our limited knowledge of the
nuclear radii should not cause any problems. In the present work we will demonstrate that it is
exactly the opposite. Even very small errors in experimental nuclear radii lead to greatly
amplified uncertainties of the King-plot nonlinearities, which bring about problems in asserting
even the order of the magnitude of the effect. This makes the King-plot nonlinearities a very
sensitive tool for studying nuclear radii, but diminish their valuableness for searches of new
physics.

{\em Arbitrary atom.---} Let us consider the isotope shift of the energy of the transition $a$
between the isotopes with the mass numbers $A_i$ and $A_0$, which will be denoted as $E_a^i$ (the
reference isotope $A_0$ will be implicit everywhere and suppressed in the notations for brevity).
It is convenient to introduce the so-called reduced isotope-shift energies $n_a^i$ as
\begin{align}
n_a^i = \frac{E_a^i}{\nicefrac{m}{M_i} - \nicefrac{m}{M_0}}\,,
\end{align}
where $m$ is the electron mass and $M_k$ is the nuclear mass of the isotope $A_k$. Within the
standard formulation, the reduced isotope-shift energy is represented as a sum of the mass-shift
and the field-shift contributions,
\begin{align}\label{eq:2}
n_a^i = K_a + F_a\, r^{(2)}_i\,,
\end{align}
where $K_a$ and $F_a$ are the mass-shift and the field-shift constants, respectively, $ r^{(2)}_i
\equiv \left[\nicefrac{R^2_i}{\lambdabar_C^2} - \nicefrac{R^2_0}{\lambdabar_C^2}\right]/
\left[\nicefrac{m}{M_i} - \nicefrac{m}{M_0}\right]$, $R_k$ is the root-mean-square (rms) nuclear
charge radius of the isotope $A_k$, and $\lambdabar_C= \nicefrac{\lambda_C}{2\pi}$ is the reduced
Compton wavelength.

It is usually assumed that the isotope-shift constants $K_a$ and $F_a$ depend only on the
transition but not on the isotope. In this case, by considering two transitions, $a$ and $b$, and
three pairs of isotopes $(A_i,A_0)$ with $i = 1$, 2, and 3, one can form the so-called King's
plot \cite{king:63}. It is easy to show that three points $(x_i,y_i) =
(n_a^1,n_b^1),(n_a^2,n_b^2),(n_a^3,n_b^3)$ lie on a straight line,
\begin{align}
y = \Big(K_b - \frac{F_b}{F_a}\,K_a\Big) + \frac{F_b}{F_a}\,x\,.
\end{align}
It is important that the linear dependence of $(n_a^i,n_b^i)$ does not rely on a theoretical
knowledge of the isotope-shift constants; the only theoretical input is the representation of the
isotope shifts in the form of Eq.~(\ref{eq:2}). This representation is remarkably accurate, but
at the level of the present-day experimental precision it may be not accurate enough.

Let us now consider a generalization of Eq.~(\ref{eq:2}) that takes into account that the
field-shift, in addition to $R^2$, depends also on a higher power of $R$. Specifically, we write
\begin{align}\label{eq:5}
n_a^i = K_a + F_a\, r^{(2)}_i + F_a w_a r^{(3)}_i \,,
\end{align}
where $ r^{(3)}_i \equiv \left[\nicefrac{R^3_i}{\lambdabar_C^3} -
\nicefrac{R^3_0}{\lambdabar_C^3}\right]/ \left[\nicefrac{m}{M_i} - \nicefrac{m}{M_0}\right]$ and
$w_a$ is an additional isotope-shift constant, which depends on the transition but not on the
isotope. We keep in mind that $r^{(3)}_i \approx (\nicefrac{R_0}{\lambdabar_C})\,r^{(2)}_i \ll
r^{(2)}_i$ since $R_0/\lambdabar_C \approx 0.01$.

Obviously, the three points $(n_a^i,n_b^i)$ of Eq.~(\ref{eq:5}) lie no longer on a straight line
but rather on a parabola, $y = a + bx + cx^2$. The nonlinearity of this function is connected
with the coefficient $c = (\nicefrac{1}{2})\,y''(x)$. Evaluating the second derivative of the
exact fit of the three points $(n_a^i,n_b^i)$ and neglecting $r^{(3)}_i$ on the background of
$r^{(2)}_i$, we arrive at the following expression
\begin{align}\label{eq:7}
y'' = 2\, \frac{F_b}{F_a^2}\,(w_a-w_b)\,{\cal P}_{\rm nucl}\,,
\end{align}
where
\begin{align} \label{eq:8}
{\cal P}_{\rm nucl} = \frac{r^{(2)}_3(r^{(3)}_1-r^{(3)}_2) + r^{(2)}_1(r^{(3)}_2-r^{(3)}_3) + r^{(2)}_2(r^{(3)}_3-r^{(3)}_1)}
  {(r^{(2)}_1-r^{(2)}_2) (r^{(2)}_1-r^{(2)}_3) (r^{(2)}_2-r^{(2)}_3)}\,.
\end{align}
It is now clear that if we would like to predict the nonlinearity of King's plot in the
approximation of Eq.~(\ref{eq:5}), we have to calculate the field-shift constants $F_{a,b}$ and
$w_{a,b}$ and multiply them by the nuclear factor ${\cal P}_{\rm nucl}$, which is determined by
the experimental nuclear parameters (charge radii and masses) of the isotopes.

Let us now address the question: to which accuracy can we determine the nuclear factor ${\cal
P}_{\rm nucl}$ basing on the available experimental data? We consider an example of four isotopes
of tin ($Z = 50$) with $A_i = (118,120,122,124)$. The nuclear radii are taken \cite{angeli:13} as
$R_{118} = 4.6393\,(1)$~fm, $R_{120} = 4.6519$~fm, $R_{122} = 4.6634\,(1)$~fm, and $R_{124} =
4.6735\,(1)$~fm. Note that we keep only the {\em relative} uncertainty of the radii with respect
to the $A = 120$ isotope, omitting the common systematic uncertainty of 0.0020~fm, which will be
ignored in the present context. The nuclear masses are taken from Ref.~\cite{wang:12}; their
uncertainties do not play any role here.

We now evaluate ${\cal P}_{\rm nucl}$ numerically, varying each of $R_i$ within their
uncertainties. The result is surprising: although the nuclear radii are supposed to be known with
a five-digit accuracy, the results for  ${\cal P}_{\rm nucl}$ may vary by several orders of
magnitude! Specifically, we obtain $|{\cal P}_{\rm nucl, min}| = 5.6\times10^{-5}$ and $|{\cal
P}_{\rm nucl, max}| = 1.3\times10^{-3}$. The reason for such striking behaviour is that both the
numerator and the denominator in Eq.~(\ref{eq:8}) can nearly vanish for some combinations of
$R_i$, leading to very small or very large values of ${\cal P}_{\rm nucl}$.

We have to conclude that regardless of the accuracy of theoretical calculations of the
isotope-shift constants, any predictions for the nonlinearities of the King plot are problematic
because of their extreme sensitivity to errors of experimental nuclear radii. A way to circumvent
this problem might be to consider the ratio of the nonlinearities for two pairs of transitions.
We note that in Eq.~(\ref{eq:7}) the nuclear part is factorized out and, therefore, the ratio of
the second derivatives $y''$ for two pairs of transitions is free from the nuclear uncertainties.
However, the exact factorization holds only for one-parameter extensions of the standard formula
like Eq.~(\ref{eq:5}).

The realistic situation is more complicated. In particular, the field-shift constant contains
terms with $\ln R$ in its expansion and is modified by the nuclear-polarization effects which
often demonstrate irregular dependence on the isotope mass number. The general expression for the
reduced isotope-shift energy can be written as
\begin{align}\label{eq:9}
n_a^i = K_{a}^i + F_{a}^i\, r^{(2)}_i \,,
\end{align}
where the isotope-shift ``constants'' $K_{a}^{i}$ and $F_{a}^{i}$ depend on the isotope
parameters (which is indicated by the superscript $i$). For a general atom, theoretical
predictions of the isotope dependence of these constants are rather complicated. In the present
work we will address this problem for the simplest case of H-like ions.

{\em H-like ions.---} We now turn to examining the case of one-electron ions. From now on, we
will adopt the relativistic units with $c = \hbar = 1$, which greatly simplifies the following
formulas. We start with the mass shift. The main isotope dependence of the mass-shift constant
comes from the quadratic part $\sim\!(\nicefrac{m}{M})^2$ of the nuclear recoil effect.
Specifically, we write
\begin{align}\label{eq:10}
K_a^i = K_a^{(1)} + \frac{m}{M_i}\, K_a^{(2)}\,,
\end{align}
where the first-order mass-shift constant $K^{(1)}$ is well-known, see, e.g., review
\cite{yerokhin:15:Hlike}, and $K^{(2)}$ is the second-order mass-shift constant calculated in the
present work. For a reference state with quantum numbers $(njl)$, the second-order mass-shift
constant is expressed as a second-order perturbation correction induced by the relativistic
recoil operator ${H}_{\rm rec}$ \cite{shabaev:85},
\begin{equation}\label{eq:10:a}
    K^{(2)}_{njl}= \sum_{k\neq (njl)}\frac{ \lbr njl | {H}_{\rm rec}| k \rbr \, \lbr k | {H}_{\rm rec}| njl \rbr }{\varepsilon_{njl}-\varepsilon_k}
    \,,
\end{equation}
where summation over $k$ is performed over the complete Dirac spectrum and
\begin{equation}
    {H}_{\rm rec}=\frac{1}{2}
    \left[\bfp^2-\frac{Z\alpha}{r}\left(\balpha+\frac{(\balpha\cdot\bfr)\,\bfr}{r^2}\right)\cdot\bfp \right]\,.
    \label{eq:H rec}
\end{equation}
Here, $\bfp$ is the electron momentum and ${\balpha}$ is the vector of Dirac matrices. We
calculate $K^{(2)}_{njl}$ in two ways: analytically, performing the summation over the Dirac
spectrum with help of generalized virial relations \cite{shabaev:91:jpb}, and numerically, with
the finite basis set $B$-spline method \cite{shabaev:04:DKB}. Results obtained by two different
methods agree with each other. For light ions, we also observe good agreement with known results
for the first terms of the $Z\alpha$ expansion \cite{yerokhin:15:Hlike}
\begin{equation}
    K^{(2)}_{njl}=-\frac{(Z\alpha)^2}{2 n^2}+\frac{(Z\alpha)^4}{2 n^3}\left(\frac{3}{2n} - \frac{2}{2 l+1}\right) + \ldots\,.
\end{equation}

The isotope dependence of the field-shift constant arises from two main sources: (i) the
deviation of the $R$-dependence of the finite nuclear size (fns) correction from the $R^2$ form
and (ii) the effect of the nuclear polarization. The fns correction to the energy levels $\Delta
E_{\rm fns}$ is calculated numerically, by solving the Dirac equation with the potential of an
extended nuclear charge, see Ref.~\cite{yerokhin:11:fns} for details. The obtained results are
represented, factorizing out the leading $R$ dependence \cite{shabaev:93:fns}, as
\begin{align}\label{eq:14}
        \Delta E_{\rm fns}&= R^{2\gamma} \, G_{\rm fns}(\Za, R)\,,
\end{align}
where $\gamma = \sqrt{(j+\nicefrac{1}{2})^2-(\Za)^2}$ and $R$ is the nuclear rms radius. The
field-shift constant is obtained from the fns correction to the transition energy as $F_a^i =
\nicefrac{\Delta E_{{\rm fns}, a}}{R^2}$. We note that the deviation of the $R^{2\gamma}$ term in
Eq.~(\ref{eq:14}) from the standard $R^2$ factor does not cause any nonlinearity of King's plot,
because it leads to a multiplication of $F$ in Eq.~(\ref{eq:2}) by a factor that does not depend
on the transition. So, the fns contribution to the nonlinearity comes only from the $R$
dependence of the function $G_{\rm fns}$. We note that this dependence is very weak; in order to
reliably calculate it for medium-$Z$ ions, we had to solve the Dirac equation in the
extended-precision arithmetics.

%%%%%%%%%%%%%%%%%%%%%%%%%%%%%%%%%%%%%%%%%%%%%%%%%%%%%%%%%%%%%%%%%%%%%%%%
%%%%%
%%%%%
%%%%%%%%%%%%%%%%%%%%%%%%%%%%%%%%%%%%%%%%%%%%%%%%%%%%%%%%%%%%%%%%%%%%%%%
\begin{figure}
\centerline{
\resizebox{.95\columnwidth}{!}{%
  \includegraphics{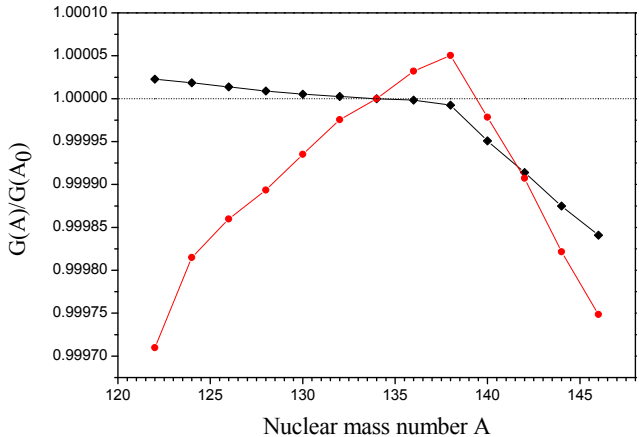}
}}
 \caption{The isotope dependence of the finite nuclear size and nuclear polarization effects, with the leading
 $R^{2\gamma}$ dependence factorized out, for the $1s$ state
 of H-like barium ($Z = 56$). Black diamonds show the finite nuclear size correction, in terms of the function
 $G_{\rm fns}$ defined by Eq.~(\ref{eq:14}). The red circles
 present results for the combined nuclear size and nuclear polarization effect, in terms of the function
 $G_{\rm nucl} = G_{\rm fns}(1-\nicefrac{1}{1000}\,g_{\rm np})$.
 Both functions $G_{\rm fns}(A)$ and $G_{\rm nucl}(A)$ are normalized to their values for the isotope $A_0 = 134$.
 \label{fig:Gfunc}}
\end{figure}

%%%%%%%%%%%%%%%%%%%%%%%%%%%%%%%%%%%%%%%%%%%%%%%%%%%%%%%%%%%%%%%%%%%%%%%%
%%%%%
%%%%%
%%%%%%%%%%%%%%%%%%%%%%%%%%%%%%%%%%%%%%%%%%%%%%%%%%%%%%%%%%%%%%%%%%%%%%%
\begin{figure}
\centerline{
\resizebox{0.95\columnwidth}{!}{%
  \includegraphics{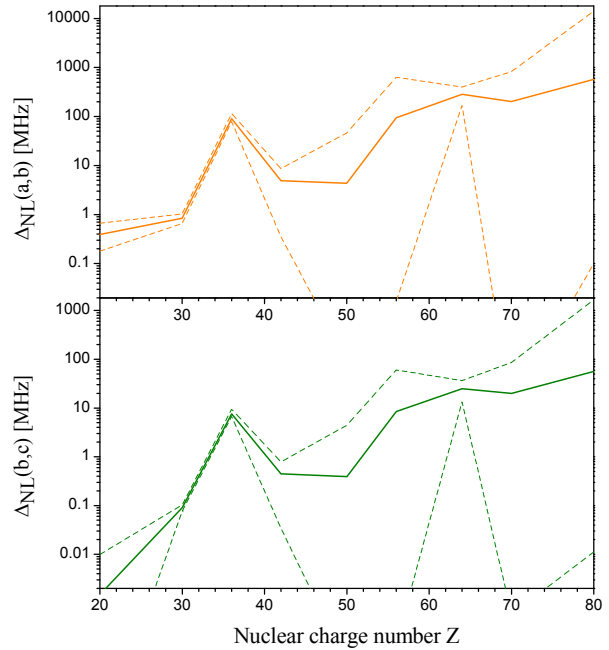}
}}
 \caption{King-plot nonlinearities of H-like ions for two transition pairs, $(a,b)$ (top) and $(b,c)$ (bottom), where
 $a = 1s \to 2p_{1/2}$, $b = 2s \to 2p_{1/2}$, and $c = 2s \to 2p_{3/2}$. The calculated points correspond to the
 isotope chains $(A_0,A_0+2,A_0+4,A_0+6)$, with the reference ($A_0$) isotopes: $^{40}$Ca, $^{64}$Zn, $^{82}$Kr, $^{92}$Mo, $^{118}$Sn,
 $^{132}$Ba, $^{154}$Gd, $^{166}$Yb, $^{196}$Hg. The dashed lines represent the variation of the nonlinearities
 when the nuclear radii are varied within their experimental uncertainties.
 \label{fig:NL}}
\end{figure}

The nuclear polarization correction to the energy of a state with quantum numbers $(njl)$ is
expressed as \cite{plunien:95,nefiodov:96}
\begin{align}\label{eq:16}
\Delta E_{\rm np} = -\alpha \sum_{LM} B(EL)\, \sum_k
\frac{ \big|\lbr njl | F_L\, Y_{LM} | k\rbr\big|^2}
 {\varepsilon_k-\varepsilon_{njl}+ {\rm sign}(\varepsilon_k)\,\omega_L}\,,
\end{align}
where $B(EL) = B(EL;L\to 0)$ are the reduced probabilities of nuclear transitions from the
excited (``$L$'') to the ground (``$0$'') level, $\omega_L$ are the nuclear excitation energies
with respect to the ground state, $F_L$ are radial functions given by Eq.~(4) of
Ref.~\cite{nefiodov:96}, $Y_{LM}$ are the spherical harmonics, and the summation over $k$ is
performed over the complete spectrum of electronic states. Multipole contributions in
Eq.~(\ref{eq:16}) are usually separated into two classes: contributions from the giant resonance
transitions and those from the lowest-lying nuclear rotational transitions. Among the latter, the
electrical quadrupole transition between the ground 0$^+$ the lowest-lying 2$^+$ state is the
dominant channel for most of even-even nuclides. Our calculation of the nuclear polarization
correction includes the dominant $E2$ nuclear rotational transition and the giant resonance
transitions with $L\le3$. Experimental results for the nuclear quadrupole transition
probabilities $B(E2)$ and excitation energies $\omega_L$ were taken from Ref.~\cite{raman:01}.
The summation over the Dirac spectrum was performed with help of the finite basis set $B$-spline
method \cite{shabaev:04:DKB}. We find good agreement with numerical results of
Ref.~\cite{nefiodov:96}. A similar calculation was recently presented in Ref.~\cite{flambaum:21};
the difference is that in that work empirical approximate formulas were used for $B(E2)$ and only
the dominant $L = 1$ giant resonance was included.

Numerical results of our calculation of the nuclear polarization correction are conveniently
parameterized in terms of the dimensionless function $g_{\rm np}$, defined as a multiplicative
factor to the fns correction,
\begin{align}\label{eq:17}
\Delta E_{\rm np} = -\frac{g_{\rm np}(Z,A)}{1000}\,\Delta E_{\rm fns}\,.
\end{align}
We studied the nuclear polarization effect for the $1s$, $2s$ and $2p_{1/2}$ states; for the
$2p_{3/2}$ state the corresponding correction is very small and thus neglected. We find that
$g_{\rm np}$ varies from $0.3$ to $1.0$ for all nuclei encountered in the present work, $Z\in
(20,92)$. It is instructive to compare the isotope dependence of the fns and the nuclear
polarization effects. Such a comparison is presented in Fig.~\ref{fig:Gfunc} for a chain of
isotopes of Ba$^{55+}$. We observe that both effects yield significant contributions, but the
nuclear polarization is larger and, more importantly, its isotope dependence is non-monotonic.
This is connected with the behaviour of the nuclear transition probabilities $B(E2)$, which tend
to grow as the nuclide mass number moves away from the stability region. We note that the kink of
the plot for $G_{\rm fns}$ is due to unevenness of the dependence of $R$ on the mass number $A$.

We are now in position to calculate the King-plot nonlinearities for H-like ions. We adopt the
definition \cite{yerokhin:20:kingsplot,flambaum:18} of the nonlinearity of a 3-point curve as a
shift of the ordinate of the third point from the straight line defined by the first two points,
\begin{equation}
    \delta \epsilon^{3}_{ab}= \Big(\frac{m}{M_3}-\frac{m}{M_0} \Big)
    \left[n^{3}_b-n^{1}_b-\frac{n^{2}_b-n^{1}_b}{n^{2}_a-n^{1}_a}\left(n^{3}_a-n^{1}_a\right)\right].
    \label{eq:18}
\end{equation}
The nonlinearity of the King plot for the transition pair $(a,b)$ is defined by symmetrizing the
above expression with respect of $a$ and $b$,
\begin{equation}
    \Delta_{\rm NL}(ab)=\nicefrac{1}{2}\left(\big|\delta \epsilon^{3}_{ab}\big|+\big|\delta \epsilon^{3}_{ba}\big|\right).
    \label{eq:19}
\end{equation}
The nonlinearity $\Delta_{\rm NL}$ has a dimension of energy and approximately indicates the
experimental accuracy needed in order to detect it. We note that the numerical evaluation of
$\Delta_{\rm NL}$  involves large numerical cancellations (due to multiple subtractions of
similar numbers) and needs to be performed in an extended-precision arithmetics.

Results of our numerical calculations of the King-plot nonlinearities for different H-like ions
are presented in Fig.~\ref{fig:NL}. We studied two pairs of transitions, $(a,b)$ and $(b,c)$,
with $a = 1s\to 2p_{1/2}$, $b = 2s\to 2p_{1/2}$, and $c = 2s\to 2p_{3/2}$. From Fig.~\ref{fig:NL}
we make several conclusions which could have been anticipated from our earlier analytical
considerations. First, we find that the calculated nonlinearities behave irregularly as a
function of $Z$, because they crucially depend on differences of charge radii of the nuclides.
Second, we find that the experimental errors of the nuclear radii cause tremendously amplified
uncertainties of the resulting nonlinearities. Generally, only the upper bounds of the
nonlinearities can be predicted and these bounds depend crucially on the assumed uncertainties of
the nuclear radii. We would like to stress that for our analysis we selected the nuclides with
the best-known charge radii. Specifically, the relative uncertainties of the rms radii for the
studied nuclides with $Z > 40$ are about 0.0001~fm \cite{angeli:13}. For less known isotopes, the
uncertainties of nonlinearities are larger by orders of magnitude.

Although our numerical calculations were performed for the specific choice of H-like ions, we
expect that the situation remains qualitatively the same for many-electron atoms, including the
case of singly charged ions which are most relevant from the experimental point of view. This
expectation is supported by the analytical example discussed in the first part of the present
paper. In order to obtain quantitative results for the upper bounds for the Standard-Model
King-plot nonlinearities for singly charged ions, dedicated calculations are needed for each
particular element.

The observed hypersensitivity of the King-plot nonlinearities on the nuclear radii can be used
for improving our knowledge of the isotope differences of the nuclear charge radii. For example,
if one of the three differences of the nuclear radii involved in the King plot is known
significantly worse than the other two, it should be possible to improve its value basing on the
measured King-plot nonlinearity.

The dependence of the predicted King-plot nonlinearities on the nuclear radii can be
significantly reduced if we study the ratio of the nonlinearities for two transition pairs, e.g.,
$\Delta_{\rm NL}(ab)/\Delta_{\rm NL}(bc)$. This can be seen from the analytical example of
 Eq.~(\ref{eq:7}) and also from
the fact that the plots for $\Delta_{\rm NL}(ab)$ and $\Delta_{\rm NL}(bc)$ in Fig.~\ref{fig:NL}
look very similar. However, the factorization of the nuclear degrees of freedom, exact in the
case of Eq.~(\ref{eq:7}), becomes only approximate in the complete calculation. As a consequence,
we still encounter combinations of nuclear radii leading to strong variations of the ratio of the
nonlinearities.

Turning to perspectives of using the King-plot nonlinearities for searches of new physics beyond
the Standard Model, we conclude that the findings of the present work significantly complicate
such searches. More specifically, when an experiment detects a nonlinearity (as, e.g., in
Refs.~\cite{miyake:19,counts:20}), it should be  very difficult to distinguish whether this is
caused by the new physics or the standard atomic theory, because of our limited knowledge of the
nuclear charge radii. Therefore, the upper bounds on the new physics derived from such an
observation will be defined by the observed nonlinearity and any further progress in experimental
precision will not improve these bounds. An alternative approach would be to use independent
constraints to eliminate the possibility of the new physics
\cite{bordag:11,raffelt:12,hardy:17,delaunay:17} and interpret the observed nonlinearities in
terms of constraints on the nuclear radii.

Summarizing, we calculated nonlinearities of the King plot originating from the Standard-Model
atomic theory. The calculations were performed first analytically for a model problem applicable
for an arbitrary atom. Next we performed a detailed numerical calculation for hydrogen-like ions,
taking into account the quadratic nuclear recoil, the higher-order finite nuclear size and the
nuclear polarization effects. We found that the Standard-Model predictions of the King-plot
nonlinearities are hypersensitive to experimental errors of nuclear charge radii, often leading
to uncertainties even in the order of magnitude of the effect. Our restricted knowledge of the
nuclear radii leads to the greatly magnified ambiguities in the Standard-Model predictions of the
nonlinearities, thus greatly complicating an identification of possible effects originating from
new physics.

{\em Acknowledgments.---} The calculations for arbitrary atoms were supported by the Russian
Science Foundation, Grant No. 20-62-46006. The calculations for H-like ions were supported by
Deutsche Forschungsgemeinschaft through Grant No. SU 658/4-1.

%\bibliographystyle{c:/-a-/papers/bibtex/phaip30}
%\bibliography{c:/-a-/papers/bibtex/hfst}

\end{document}